\documentclass[sigconf]{acmart}

\AtBeginDocument{%
  }

\copyrightyear{2024}
\acmYear{2024}
\setcopyright{rightsretained}
\acmConference[UIST Adjunct '24]{The 37th Annual ACM Symposium on User
Interface Software and Technology}{October 13--16, 2024}{Pittsburgh, PA, USA}
\acmBooktitle{The 37th Annual ACM Symposium on User Interface Software and
Technology (UIST Adjunct '24), October 13--16, 2024, Pittsburgh, PA, USA}
\acmDOI{10.1145/3672539.3686345}
\acmISBN{979-8-4007-0718-6/24/10}

\begin{document}

\title{Efficient Optimal Mouse Sensor Position Estimation using Simulated Cursor Trajectories}

\author{Minhyeok Baek}
\email{minheak06@dgist.ac.kr}
\orcid{0000-0002-5789-2033}
\affiliation{%
  \institution{Daegu Gyeongbuk Institute of Science and Technology}
  \city{Daegu}
  \country{Republic of Korea}
}

\author{Sunjun Kim}
\authornote{Correspondence author}
\email{sunjun_kim@dgist.ac.kr}
\affiliation{%
  \institution{Daegu Gyeongbuk Institute of Science and Technology}
  \city{Daegu}
  \country{Republic of Korea}
}

\renewcommand{\shortauthors}{Baek et al.}

\begin{abstract}
The optimal sensor position on a computer mouse can improve pointing performance, but existing calibration methods require time-consuming repetitions of pointing tasks. In this paper, we propose a novel calibration approach that dramatically reduces the time and effort required to determine a user's optimal mouse sensor position. Our method simulates cursor trajectories for different sensor positions using a dual-sensor mouse, eliminating the need for repetitive measurements with multiple sensor placements. By analyzing the straightness of the simulated paths, quantified by the mean absolute error (MAE) relative to an ideal straight-line path, we estimate the sensor position that would yield the most efficient pointing motion for the user. Our preliminary results indicate that the proposed simulation-based calibration method could reduce the calibration time from an hour to just five minutes, while providing a better identification of the optimal mouse sensor positions.
\end{abstract}



\maketitle

\section{Introduction}
The position of a computer mouse sensor could change the pointing performance \citep{kim2020optimal, verplank1989microsoft}, and the optimal position of the sensor varies between individuals and can improve the pointing performance.

There has been limited research on the impact of mouse sensor placement. 
The first academic study, conducted by Verplank and Oliver \citep{verplank1989microsoft} in 1989, explored three early mouse prototypes. They found that the mouse with the sensor at the front exhibited better performance, but their study had methodological limitations and used outdated technology. 

More recently, Kim et al. \citep{kim2020optimal} investigated the effect of sensor position using modern optical mice.
The study indicated that altering the sensor position results in varied cursor movements. Also, each individual had a different preference for optimal sensor placement.
The sensor placed in the center is generally preferred, and further personalized sensor position can increase the pointing performance by 4\% on average.
However, their calibration process required time-consuming Fitts' law \citep{mackenzie1992fitts} throughput measurements at multiple sensor positions, which took approximately an hour to complete.

In this study, we introduce a novel calibration method that dramatically reduces the time and effort needed to find the ideal mouse sensor position for a user. 
Our research extends the findings of Kim et al. \citep{kim2020optimal} by proposing a more efficient calibration approach utilizing the benefit of the dual sensor mouse.
The new method only takes about five minutes, from which all sensor positions are simulated and evaluated post hoc after data collection. This is a substantial reduction compared to the original hour-long method.

\section{Virtual Mouse Sensor Position}
We used a dual sensor mouse \cite{kim2020optimal}, which captures displacements from two sensors: front \((dX_{front}, dY_{front})\) and rear \((dX_{rear}, dY_{rear})\). The cursor position \((mX, mY)\) for a virtual sensor at a position \(p\) (where \(p\) is between 0.0 and 1.0) is calculated as:

\[ mX = k \left[ (1 - p) dX_{front} + p \cdot dX_{rear} \right] \]
\[ mY = k \left[ (1 - p) dY_{front} + p \cdot dY_{rear} \right] \]

Here, \(k\) is a gain factor that scales the combined displacement values. This allows simulation of any sensor position along the line connecting the front and rear sensors.
The question we are solving here is: \textbf{\textit{What is the best sensor position $p$ for a user?}}

\section{Fuzzy rail cursor}
An aimed pointing movement consists of multiple submovements \cite{meyer1988optimality}. 
For ideal pointing performance, the cursor should reach the target with the first submovement, and the trajectory should be as straight as possible.

\begin{figure}[h]
    \begin{center}
    \includegraphics[width=\linewidth]{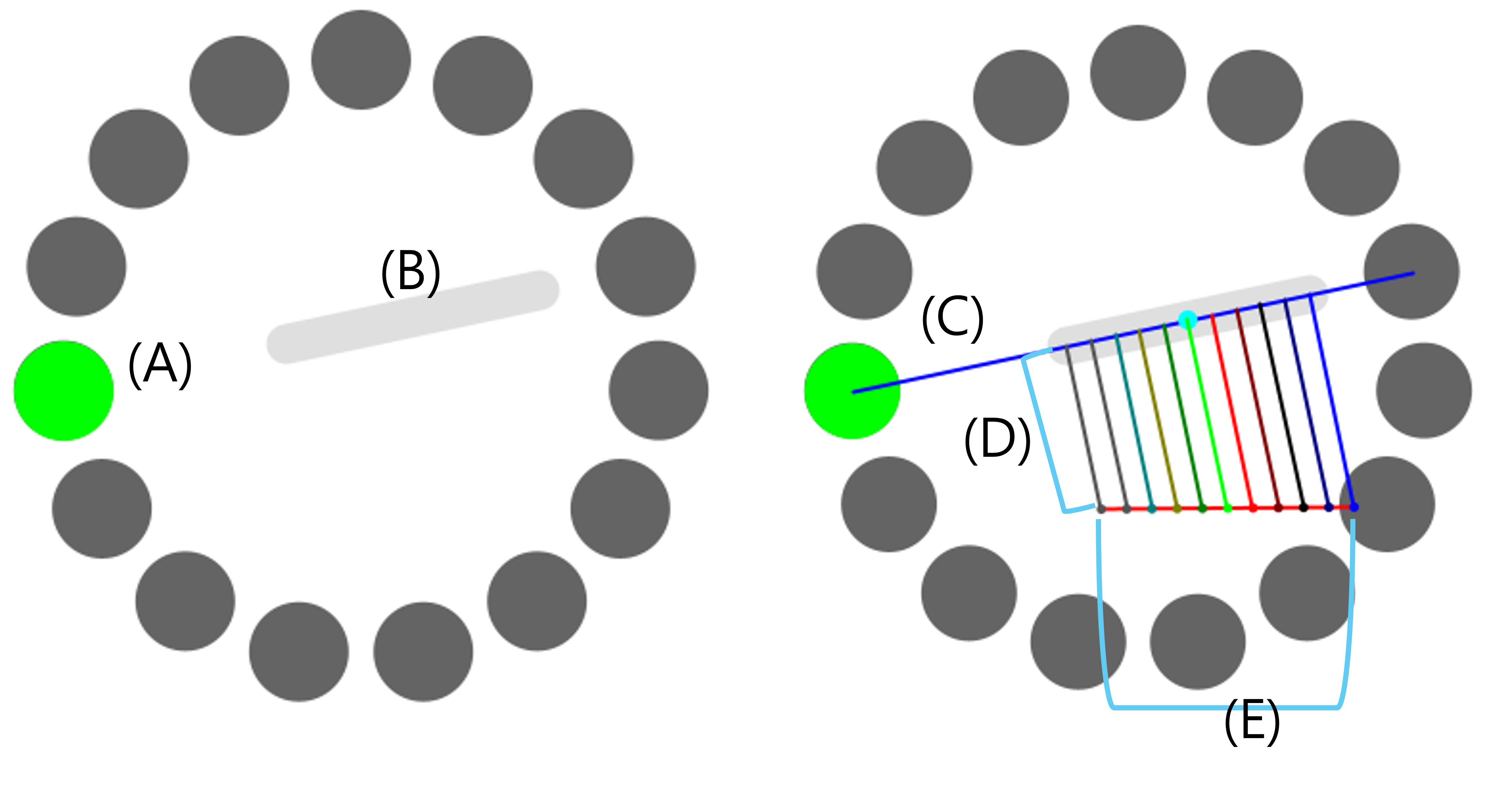}
    \caption{Left: actual rendered cursor, Right: its working principle. (A) target (B) fuzzy rail cursor (C) ideal path line, called \textit{rail} (D) projection of fuzzy cursor on the ideal line (E) fuzzy cursor from real cursor movement.}
    \label{fig:rail_method}
    \end{center}
\end{figure}

\textit{Fuzzy rail} cursor is designed with the aim of emulating the perfect scenario: hitting the target in a straight trajectory with a single submovement. 
When a point cursor is manipulated from a particular sensor position, users often adjust the deviation of the path during the pointing task. This visual feedback might cause a significant bias towards the present sensor position being used.
To eliminate this bias, it is necessary to visualize all possible cursor positions from a variety of sensor positions (from $p=0.0$ to $1.0$) simultaneously. Thus, a \textit{fuzzy} cursor was created to replace the single-point cursor (see Figure \ref{fig:rail_method}). 
Additionally, we implemented the fuzzy cursor to appear along a rail -- called \textit{fuzzy rail} cursor -- linking the starting point and the target, giving users the perception of moving towards the target in a direct cursor path. 
Finally, to achieve the pointing action in a single submovement, the cursor gradually fades starting from the midpoint and completely vanishes by the 3/4 point on the rail, thereby removing the need for additional corrective submovements.

For the calibration task, we adopted Fitts' law task (ISO 9241-411) design with an addition of the fuzzy rail cursor.
We anticipated that this cursor design would lead users to perceive the cursor as \textit{snapping} to the target, completing their target selection trial with only their initial primary submovement.

A user performs the task for about five minutes. During the task, raw sensor values\((dX_{front}, dY_{front}, dX_{rear}, dY_{rear})\) were recorded. After completion of the task, the simulated cursor trajectories were calculated while varying the position of the cursor (=$p$). Then, the Mean Absolute Error (MAE) of the trajectories was measured against the straight line (=rail).
Finally, the optimal sensor position is determined as the position with minimal MAE.

\section{Experiment and Results}
We conducted a preliminary experiment to examine the efficacy of our proposed calibration method. The previous manual calibration method \citep{kim2020optimal} was also performed as a baseline.

\begin{figure}[h]
    \begin{center}
    \includegraphics[width=\linewidth]{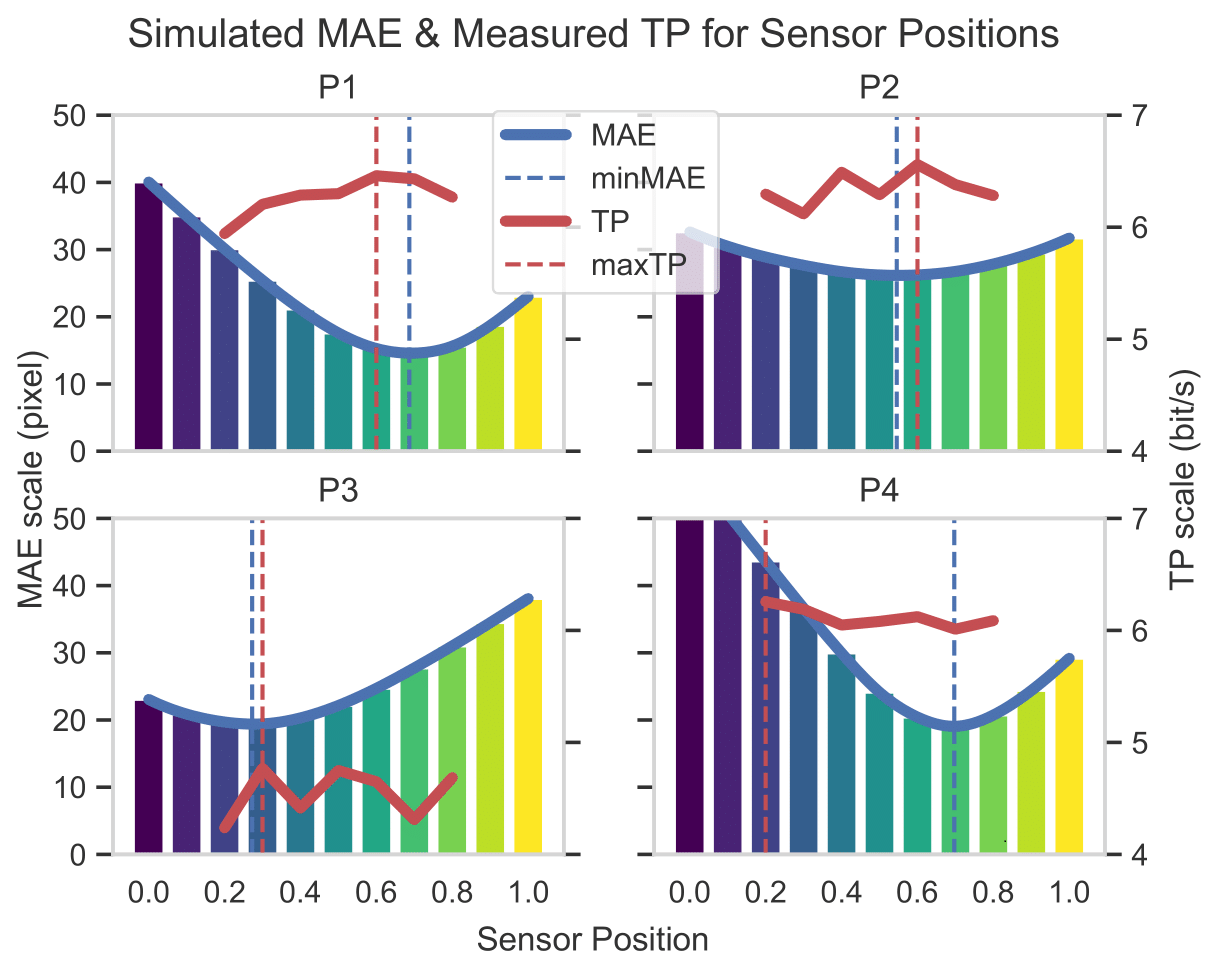}
    \caption{Averaged MAE measured in our method (blue), and averaged TP measured in manual method (red). The bar graphs indicate the MAE in 11 simulated sensor positions, and the red line indicates the TPs meausred in 7 sensor positions. The vertical line represents the best sensor position obtained from the two methods.}
    \label{fig:Result}
    \end{center}
\end{figure}

For each participant, MAE was measured using the fuzzy rail cursor method and Fitts' law throughput (TP) was measured at different sensor positions using the baseline method (Figure \ref{fig:Result}).
Except for P4, three participants (P1, P2, P3) exhibited well-aligned optimal sensor positions obtained from the two methods.

In general, our method was comparable to the result of the manual method (see Figure \ref{fig:Result}), while it only took a fraction of the time. 
This benefit will enable a greater number of mouse users to find their ideal sensor position, tailored to their needs.

One notable observation is that our method (MAE) yielded an extremely smooth curve with a prominent apex point. 
In particular, our approach could better distinguish the difference between sensor positions. For example, while P4 did not show a significant change in pointing performance using the manual method (TP), our method (MAE) clearly identified a pronounced peak at the position $p=0.7$.
In the manual method, users may need a longer period to adapt their muscle memory to each sensor position. The manual method (TP) tends to fluctuate and be noisy in all the sensor positions tested. 

\section{Conclusion and Future Work}
In this study, we introduced a novel calibration method to determine the optimal sensor position on a computer mouse, significantly reducing the time and effort involved compared to previous approaches. Our preliminary results indicate that the proposed simulation-based calibration method is not only faster but also provides a better identification of the optimal sensor position.

In future work, an immediate next step is to validate the proposed method on a larger scale. The similarity between the results of the two methods could potentially be confirmed with additional statistical tests.
Furthermore, we aim to continuously monitor the changes in the optimal sensor position over time longitudinally. 
In addition, further exploration will be conducted on how the optimal sensor position correlates with various mouse-related factors, including the shape of the mouse, the grip style, the control-display ratio, the types of task, and the user's hand and arm properties.
\begin{acks}
This work was supported by the National Research Foundation of Korea(NRF) grant funded by the Korea government(MSIT) (RS-2023-00211872 and RS-2023-00223062).
\end{acks}

\bibliographystyle{ACM-Reference-Format}
\bibliography{sample-base}


\end{document}